\documentclass{iopart}
\usepackage{amssymb}
\usepackage[dvips]{graphicx}

\begin{document}

\title[Source Tracking for Sco X-1]{Source Tracking for Sco X-1}

\author{K Hayama$^1$, S~D Mohanty$^1$, S Desai$^2$, M Rakhmanov$^1$, \\
T Summerscales$^3$, S Yoshida$^4$}
\address{$^1$The University of Texas at Brownsville, Brownsville, TX, 78520, US}
\address{$^2$ Center for Gravitational Wave Physics, the Pennsylvania State University, University Park, PA 16802, US}
\address{$^3$Department of Physics, Andrews University, Berrien Springs, MI 49104, US}
\address{$^4$Southeastern Louisiana University, Hammond, LA 70402, US}

\ead{kazuhiro.hayama@ligo.org}

\begin{abstract}
Sco X-1, the brightest low mass X-ray binary, is likely to be a source for gravitational 
wave emission. In one mechanism, emission of a gravitational wave arrests the 
increase in spin frequency due to the accretion torque in a low mass X-ray binary. Since the gravitational waveform is unknown, a detection method assuming no apriori knowledge of the signal is preferable. In this paper, we propose to search for a gravitational wave from Sco X-1 using a {{\it source tracking}} method based on a coherent network analysis. In the method, we combine data from several interferometric gravitational wave detectors taking into account of the direction to Sco X-1, and reconstruct two polarization waveforms at the location of Sco X-1 in the sky as Sco X-1 is moving. The source tracking method opens up the possibility of searching for a wide variety of signals. We perform Monte Carlo simulations and show results for bursts, modeled, short duration periodic sources using a simple excess power and a matched filter method on the reconstructed signals. 
\end{abstract}

%\pacs{04.80.Nn,07.05.Kf,95.55.Ym,95.85.Nv,95.85.Sz,97.10.-q,97.60.Gb,97.60.Jd,97.80.Jp}

\section{Introduction}
 Measured spin frequencies of neutron stars in Low Mass X-ray Binaries (LMXBs) show that there is a limiting mechanism on the increase in spin frequency due to accretion torque. One possible limiting mechanism is emission of gravitational waves~\cite{Bildsten:1998}. Then LMXBs such as Sco X-1, which is the strongest X-ray source in the sky, are candidate sources of detectable gravitational waves. Also, quasi-periodic oscillations (QPOs) have been observed in many LMXBs. Although the mechanism responsible for the QPOs is still unknown, there is a model which predicts emission of a gravitational wave associated with the QPOs~\cite{Jernigan:2001}. The Rossi X-ray timing explorer (RXTE) satellite has millisecond resolution which enables us to probe the physics of transients on rapid timescales~\cite{1996A&AS..120C.641G}. It observed QPOs in Sco X-1 at a time LIGO was operational. 

Most current models of these events assume that the gravitational wave from Sco X-1 is a continuous periodic signal. Searches for a continuous periodic gravitational wave from Sco X-1 were conducted with LIGO S2 data~\cite{S2LSCpulsar:2007}, and S4 data~\cite{S4stochastic:2007}. The two searches used different approaches: a coherent matched filter method~\cite{S2LSCpulsar:2007} and a cross-correlation method~\cite{S4stochastic:2007}. It is also possible that the nature of a gravitational wave from Sco X-1 is different from a continuous periodic gravitational wave. 

In this paper, we adopt a {\it source tracking} approach for unmodeled sources based on the coherent network analysis {\tt RIDGE}~\cite{Hayama:2007} and propose to monitor Sco X-1 for possible gravitational wave emission. In {\tt RIDGE}, data from several gravitational wave detectors are combined coherently, taking into account the antenna patterns, locations of the detectors and the sky direction to the source. {\tt RIDGE} reconstructs $h_+$ and $h_\times$ time series by solving an inverse problem of the response matrix of a detector network. Then we do source tracking by reconstructing the $h_+$ and $h_\times$ time series at the location of Sco X-1 in the sky as Sco X-1 is moving. We then apply different algorithms to the reconstructed $h_+$ and $h_\times$ time series.
We perform Monte Carlo simulations and show results on the detectability of unmodeled bursts and long, quasi-monochromatic gravitational wave signals from Sco X-1, using an excess power and a template based search.

\section{Implementation of the source tracking method}
We implement the source tracking of Sco X-1 using a regularized coherent network method {\tt RIDGE} described in~\cite{Hayama:2007}. The {\tt RIDGE} pipeline consists of two main components: data conditioning and the generation of detection statistics. The aim of the data conditioning is to whiten the data to remove frequency dependence and any instrumental artifacts in the data. 
In {\tt RIDGE}, the data are whitened by estimating the noise floor using a running median~\cite{Mukherjee:2003}. 
By using a  running median, the estimation of the noise floor can avoid the influence of a large outlier due to a strong sinusoidal signal. The whitening filters are implemented using digital finite impulse response filters with 
the transfer function $|T(f)| = 1/\sqrt{S(f)}$, where $S(f)$ is the power spectral density (PSD) of detector noise as a function of frequency $f$. The filter coefficients are obtained by using the PSD obtained from a user-specified training data segment. The digital filter is then applied to a longer stretch of data. 
Narrow-band noise artifacts known as {\em lines} are a typical feature of the output of interferometric gravitational wave detectors. Lines originate from effects associated with the functioning of the detectors such as mirror/suspension resonant modes and power line interference, calibration lines, vacuum pumps etc~\cite{Chassande-Mottin:2005}. These lines dominate over a specific band, which reduces the signal-to-noise ratio and the  accuracy in  recovering  gravitational wave signals. 
In {\tt RIDGE}, the whitening step above is followed by a line estimation and removal method by using a technique
 called {\em median based line tracker} ({\tt MBLT}, for short) described in ~\cite{Mohanty:2002}. Essentially, 
the MBLT method consists of estimating the amplitude and phase modulation of a line feature at a given carrier 
frequency. The advantage of this method is that the line removal does not affect the burst signals in any significant 
way. This is an inbuilt feature of {\tt MBLT} which uses the running median for estimation of the line amplitude and phase functions. Transient signals appear as outliers in the amplitude/phase time series and are rejected by the 
running median estimate. The resulting conditioned data is passed on to the next step, which consists of the generation of a detection statistic. The basic algorithm implemented for coherent network analysis in {\tt RIDGE} is Tikhonov regularized maximum likelihood, described in~\cite{Rakhmanov:2006}. Recent studies~\cite{Klimenko+etal:2005,Rakhmanov:2006,Mohanty+etal:2006} show the inverse problem of a response matrix of a detector network becomes an ill-posed one and the resulting variance of the solution become large. This comes from the rank deficiency of the 
detector response matrix. The amount of  rank deficiency depends on the sky location, and therefore,  plus or cross-polarized gravitational wave signals from some directions on the sky become too noisy.  In {\tt RIDGE}, we reduce this ill-posed problem by applying the Tikhonov regulator which is a function of the sky location. The input to the algorithm is a set of equal length, conditioned data segments from the detectors in a given network. The output is $h_+$ and $h_\times$ time series at the sky location of Sco X-1 in the sky, which are reconstructed as Sco X-1 is moving. Concatenating the segments of the reconstructed $h_+$ and $h_\times$ time series, one can monitor Sco X-1 for bursts or other types of gravitational wave signals. In the $h_+$ and $h_\times$ time series, signals from the direction of interest are enhanced compared with ones from other parts of the sky due to the application of the principle of maximum likelihood.

%The output, for a given sky location $\theta$ and $\phi$, is the value of the likelihood of the data maximized over all possible $h_+$ and $h_\times$ waveforms with durations less than or equal to the data segments. The maximum  likelihood values are obtained as a function of $\theta$ and $\phi$ -- this two dimensional output called a {\em skymap}. Using the entire sky map, we construct ``radial distance'' which scales the maximum likelihood by the mean location of the same quantities in the absence of a signal. Detailed description about the detection statistic is given in~\cite{Hayama:2007}.
%In case of the source tracking of a known source, we can use the information about the sky location of the source. The information about the sky location reduces the degree of the parameter space of the coherent network analysis, and the regulator reduces the variance. 

%%%%%%%%%%%%%%%%%%%%
%%%%%%%%%%%%%%%%%%%%
\section{Simulations}
\subsection{Information about Sco X-1}
\begin{figure}
\begin{center}
\includegraphics[scale=0.5]{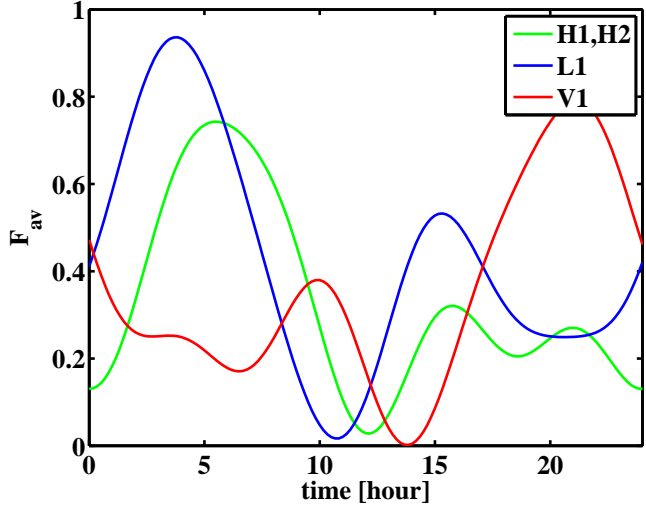}
\caption{Averaged antenna pattern $F_{\rm av}$ to the direction of the Sco X-1. Zero time corresponds to 11:00am (Sep. 12, 2007 (UTC)). H1, H2 are the 4~km, 2~km LIGO Hanford detector with the same location and orientation, L1 is the LIGO Livingston detector, and V1 is the VIRGO detector.
\label{fig:detresp_scox1}
}
\end{center}
\end{figure}

Sco X-1 is the strongest X-ray emitting LMXB on the sky. The right ascension (RA) is 16h~19m~55.085s, declination (DEC) is -15$^\circ$~38'24.9'', and the distance from the earth is $2.8\pm0.3$~kpc~\cite{BradshawFomalontGeldzahler:1999}. 
Figure~\ref{fig:detresp_scox1} shows the averaged detector antenna pattern ($F_{\rm av}$) of LIGO and VIRGO to the location of Sco X-1 as a function of time since GPS time 873630000 (11:00am, Sep. 12, 2007 (UTC)). $F_{\rm av}$ is defined as $F_+^2+F_\times^2$, where $F_{+}$ and $F_\times$ are the antenna pattern function of a plus and cross polarized gravitational wave. From this figure, for the LIGO only network, the value of the 
squared antenna pattern exceeds 0.5 for about 33\% of the time during a day. If VIRGO is added to this network, this network coverage increases to  63\%, and the detection efficiency is much improved~\cite{Hayama:2008}. Thus the LIGO-VIRGO network is quite effective to monitoring Sco X-1 compared with the LIGO only network. 

\subsection{Detection of unmodeled bursts}
\begin{figure}
\begin{center}
\includegraphics[scale=0.5]{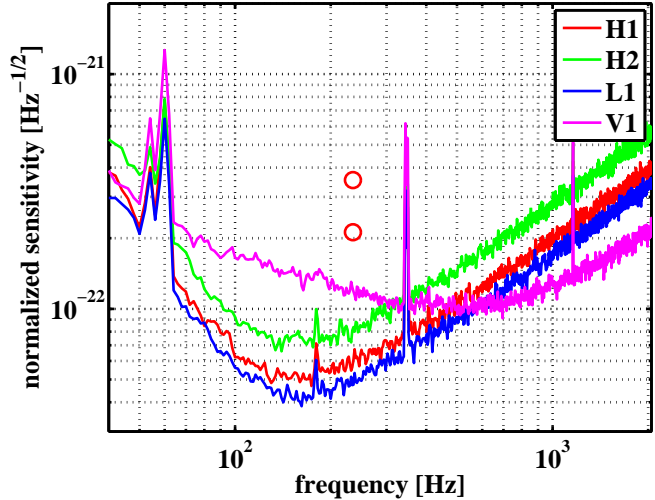}
\caption{The normalized sensitivity
%, which is defined as the amplitude spectral density of simulated H1, H2, L1, V1 data divided by $F_{\rm av}^{1/2}$,
 and root-sum-square of h of the injected sine Gaussian signals ($Q=9$) used in the Monte Carlo simulations. The open circles are the injected signals of $h_{\rm rss}= [2.1, 3.5] \times 10^{-22} {\rm Hz}^{-1/2}$ with the center frequencies 235~Hz.
\label{fig:simpsds}
}
\end{center}
\end{figure}
We performed Monte Carlo simulations to estimate the detection efficiency for Sco X-1. The network consisted
of the 4~km and 2~km LIGO Hanford (for short H1, H2), LIGO Livingston (L1) and VIRGO (V1). For the detector noise amplitude spectral densities, we used the design sensitivity curves for the LIGO and VIRGO detectors as given in~\cite{LIGOdesignsens}~\cite{VIRGOdesignsens} and we kept the locations and orientations the same as the 
real detectors. For H2, the sensitivity is $\sqrt{2}$ times less than H1. Gaussian, stationary noise was generated
 ($\sim1000$~seconds) by first generating 4 independent realizations of white noise and then passing them through finite impulse response (FIR) filters having transfer functions that {\em approximately} match the design curves. 
To simulate instrumental artifacts, we added sinusoids with large amplitudes at $(54, 60, 120, 180, 344, 349, 407, 1051)$~Hz. 

Signals with a fixed amplitude were added to the simulated noise at regular intervals. The injected signals corresponded to a single source located at
${\rm RA} = 16.3~{\rm hours}$ and ${\rm DEC} = -15.8~{\rm degrees}$, which was approximately the sky location of Sco X-1. We assumed that $h_+(Q,f_c,t) = A\exp(-(2\pi f_ct)^2/2Q^2)\sin(2\pi f_ct)$
and $h_\times (Q,f_c,t) = A\exp(-(2\pi f_ct)^2/2Q^2)\cos(2\pi f_ct)$, where $t$ is time, $Q=9$, was the Q-value and $f_c$ was the central frequency. The signal strength, $A$, was specified in terms of root-sum-square defined as $h_{\rm rss} = \left[\int_{-\infty}^{\infty} dt\; \left(h_+^2(t) + h_\times^2(t)\right) \right]^{1/2}$. In this simulation we took $h_{\rm rss} = [2.1, 3.5] \times 10^{-22}{\rm Hz}^{-1/2}$. 
Figure~\ref{fig:simpsds} shows the averaged sensitivity, defined as the amplitude spectral density of simulated H1, H2, L1, V1 data divided by $F_{\rm av}^{1/2}$, and the injected signals. Open circles are the injected signals. The x-axis for the circles represents the center frequencies and the y-axis represents the $h_{\rm rss}$ of the signals.

\begin{figure}
\begin{center}
\includegraphics[scale=0.5]{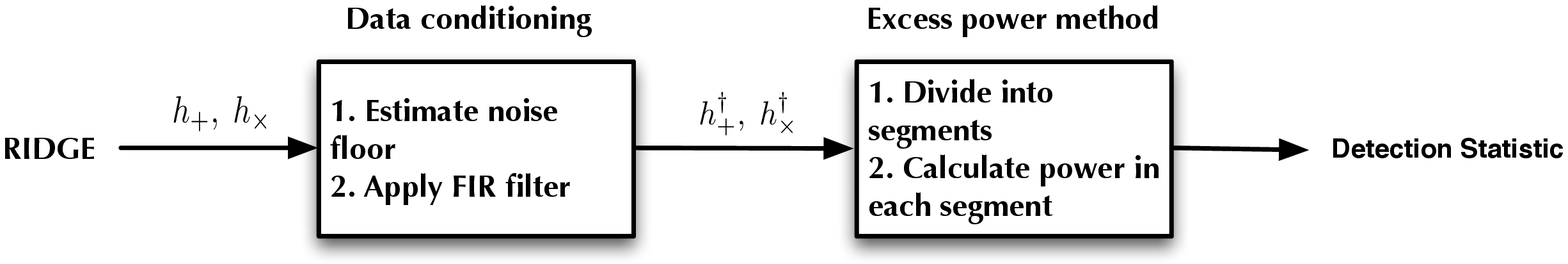}
\caption{
Block diagram for an excess power method on reconstructed $h_+$ and $h_{\times}$ time series. The reconstructed $h_+$ and $h_\times$ time series are contaminated with frequency dependent noise. They are passed on to the step of whitening. The filter coefficients are estimated using  2~seconds of data which do not contain injected signals. The whitened, reconstructed $h_+$ ($h_+^\dagger$) and $h_\times$ ($h_\times^\dagger$) are divided into 20~ms segments. Then by calculating power, one obtains detection statistics.
\label{fig:ExcessPowerFC}
}
\end{center}
\end{figure}
\begin{figure}
\begin{center}
\includegraphics[width=0.7\linewidth]{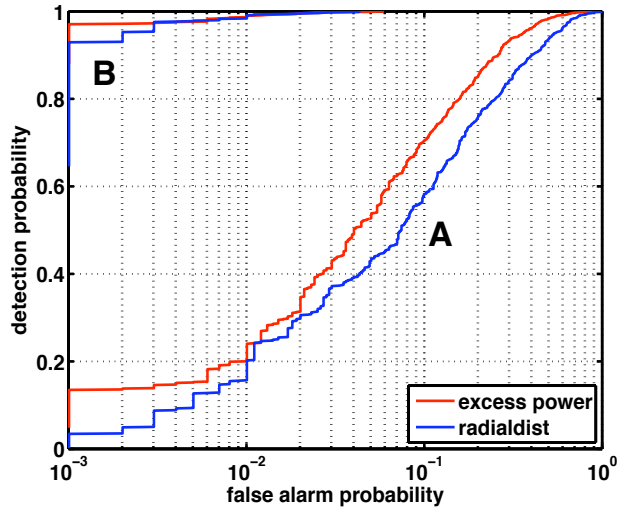}
\caption{
Receiver Operating Characteristic. Signal strength in group A is $h_{\rm rss}=2.1\times10^{-22}{\rm Hz}^{-1/2}$, and in group B is $h_{\rm rss}=3.5\times10^{-22}{\rm Hz}^{-1/2}$. Blue lines correspond to a radial distance statistic, and red lines a excess power statistic. 
\label{fig:ROCexcessvsradial}
}
\end{center}
\end{figure}

The resulting simulated data was passed on to the {\tt RIDGE} pipeline, and $h_+$ and $h_\times$ were reconstructed. We ran an excess power algorithm~\cite{Anderson:2001} on the reconstructed $h_+$ and $h_\times$. The excess power algorithm is a method to find data segments for which the power exceeds a given threshold. Figure~\ref{fig:ExcessPowerFC} shows a block diagram for the excess power method. The reconstructed $h_+$ and $h_\times$ are
 contaminated with frequency dependent noise originating from the detector. First they were whitened using the whitening filter implemented in {\tt RIDGE}. Filter coefficients were estimated using 2~second data intervals which did not contain the injected signals. The whitened reconstructed signals $h_+^\dagger$ and $h_\times^\dagger$ were divided into segments 20~ms in length. Power, which was defined as the summation over the squared amplitude of each segment, was calculated. Detection candidates were selected as segments with power beyond a given power threshold. We calculated a receiver operating characteristic (ROC) by changing values of the threshold. Figure~\ref{fig:ROCexcessvsradial} shows the ROC curves for the injected signals with $h_{\rm rss}=[2.1,\; 3.5]\times10^{-22}{\rm Hz}^{-1/2}$. For comparison, we calculated a detection statistic {\em radial distance} obtained as follows: In {\tt RIDGE} we calculated a value of the likelihood of the data maximized over all possible $h_+$ and $h_\times$ waveforms with durations less than or equal to the data segments. The maximum  likelihood values were obtained as a function of $\theta$ and $\phi$ -- this two dimensional output is called a {\em sky-map}. Using the entire sky-map, we constructed a detection statistic {\em radial distance} which scales the maximum likelihood by the mean location of the same quantities in the absence of a signal. 
ROC curves were included which were calculated using the {\em radial distance}.  ROC curves for the radial distance statistic were calculated using segments of 0.5~sec in length. This result shows even a simple excess power statistic can achieve good performance. 

\subsection{Detection of modeled bursts}
\begin{figure}
\begin{center}
\includegraphics[scale=0.5]{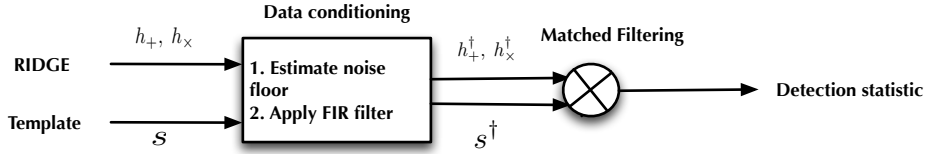}
\caption{
Block diagram for a matched filter method on reconstructed $h_+$ and $h_\times$. The reconstructed $h_+$ and $h_\times$ are passed on a whitening step. The filter coefficients are estimated using 2~seconds of data which does not contain injected signals. The template signal is also whitened by filters with the same filter coefficients. A matched filter method is then applied to the whitened, reconstructed signals.
\label{fig:flowchart_matchedfilter}
}
\end{center}
\end{figure}
We considered the detection of well-modeled bursts using a matched filter method. We used sine Gaussian signals ($Q=9$) with the center frequency of 235~Hz, $h_{\rm rss}$ of $7\times10^{-22}\rm{Hz}^{-1/2}$ as a well-modeled burst.
Figure~\ref{fig:flowchart_matchedfilter} shows a block diagram of the matched filter method we used. The reconstructed $h_+$ and $h_\times$ were whitened with the whitening filter. The filter parameters were estimated using the preceding 2~seconds of data which did not contain signals. The template was also whitened by the same whitening filter and filter parameters. The whitened data was passed on to the matched filtering defined as $T = x^Ts^{\dagger}/\sigma\sqrt{{s^\dagger}^Ts^\dagger}$,
where $x=h_{*}^\dagger\;(* =[+,\;\times])$ and $s^\dagger$ were column vectors of the whitened reconstructed polarized waveforms and the whitened template. $\sigma$ was the standard deviation of the whitened data. $T$ was the signal-to-noise ratio (SNR).

\begin{figure}
\begin{center}
\includegraphics[scale=0.3]{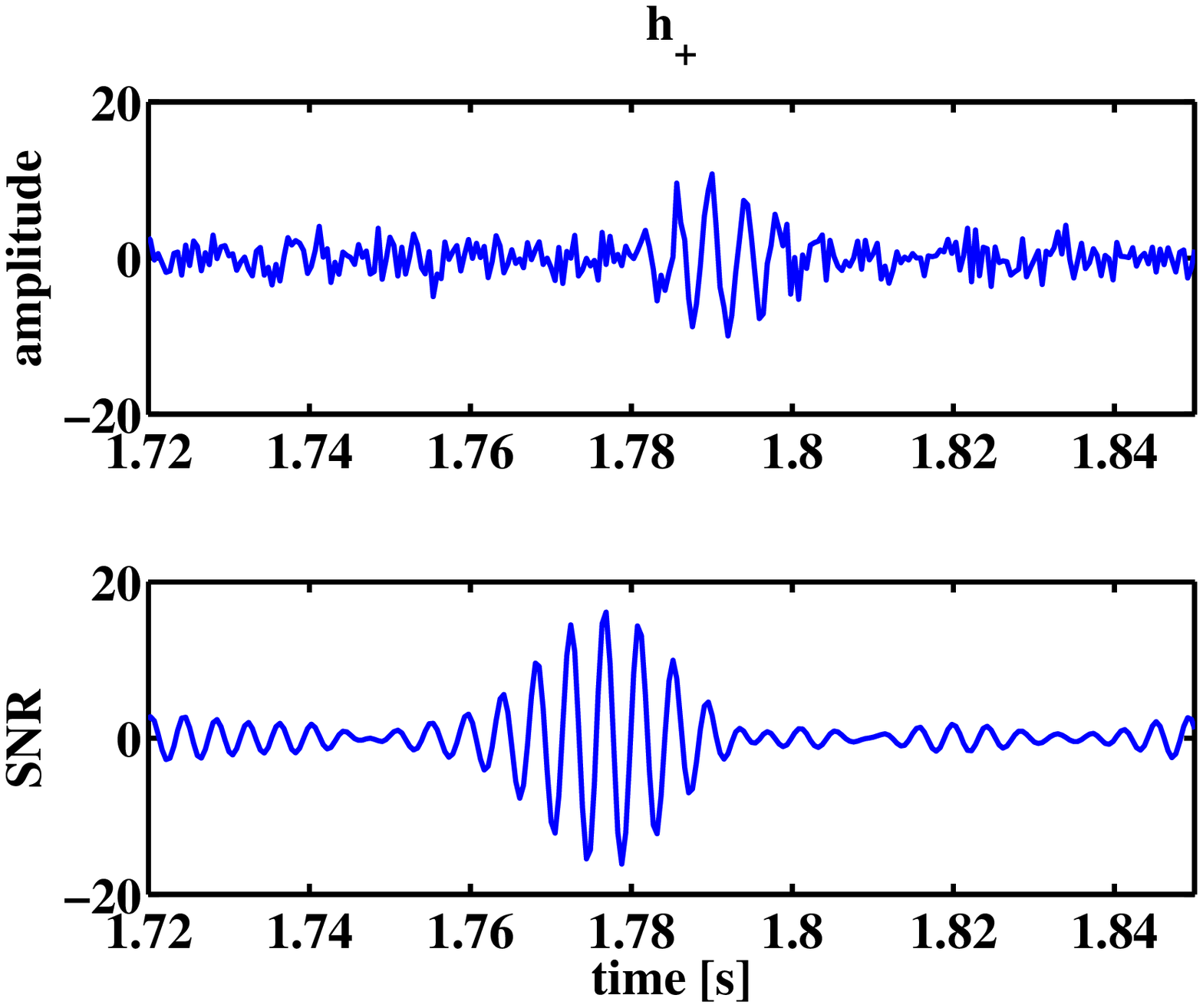}
\includegraphics[scale=0.3]{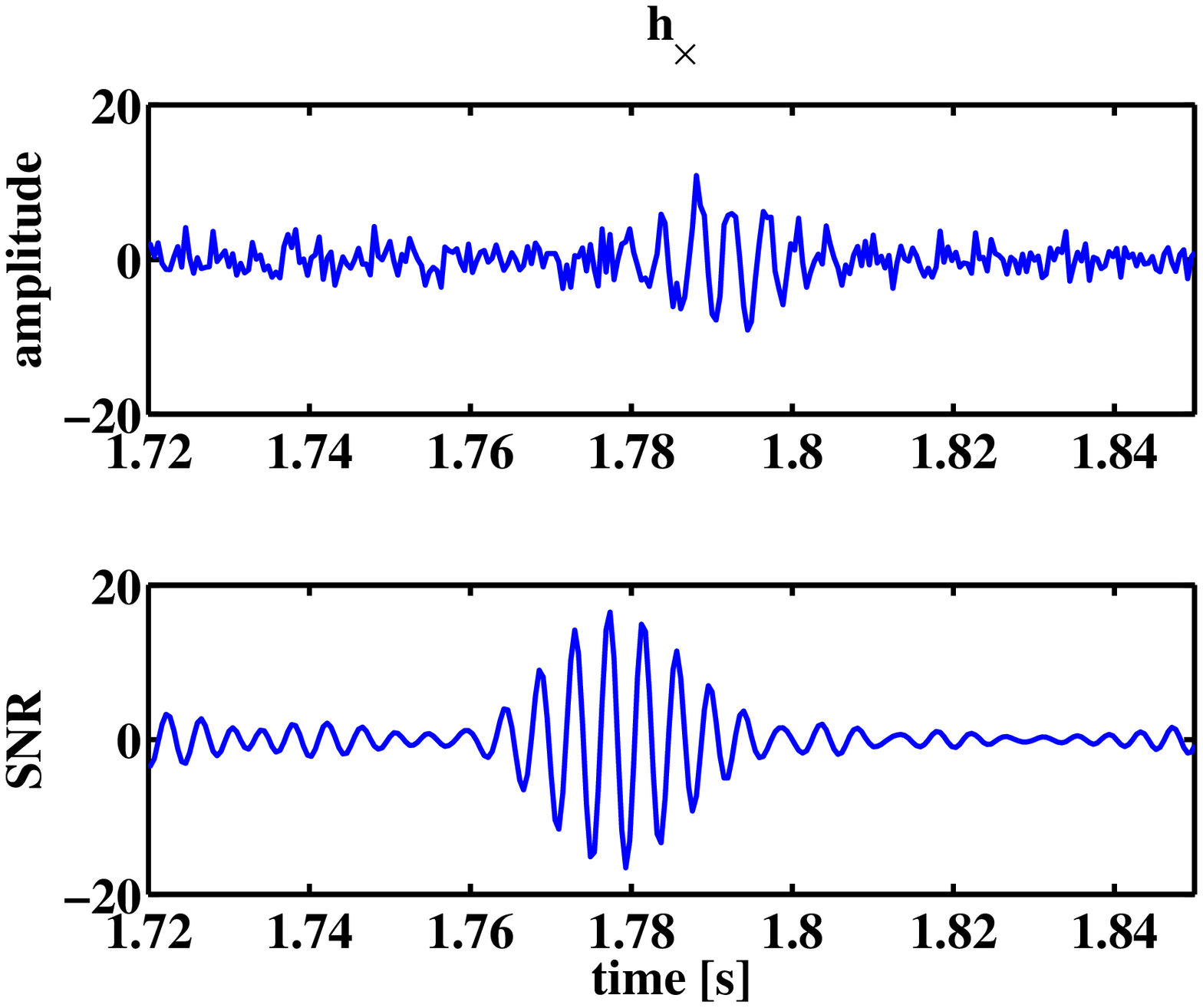}

\caption{Top plots show the reconstructed $h_+$ (left) and $h_\times$ (right). The corresponding injected signal is a sine Gaussian signal ($Q=9$) with the center frequency of 235~Hz, $h_{\rm rss}=3.5\times 10^{-22}{\rm Hz}^{-1/2}$. Bottom plots are output of the matched filter method.}
\end{center}
\end{figure}

Figure~\ref{fig:matchedfilter} is the result. The top plots show the reconstructed $h_+$ (left) and $h_\times$ (right). The bottom plots are the SNR. These plots show the signals are detected with ${\rm SNR}\simeq 15, 15$ for $h_+$ and $h_\times$ . Considering SNRs of the injected signal in H1, H2, L1, V1 are 15.2, 10.7, 18.9, 18.5 respectively, this result is reasonable.

\subsection{Detection of a monochromatic signal}
\begin{figure}
\begin{center}
\includegraphics[scale=0.3]{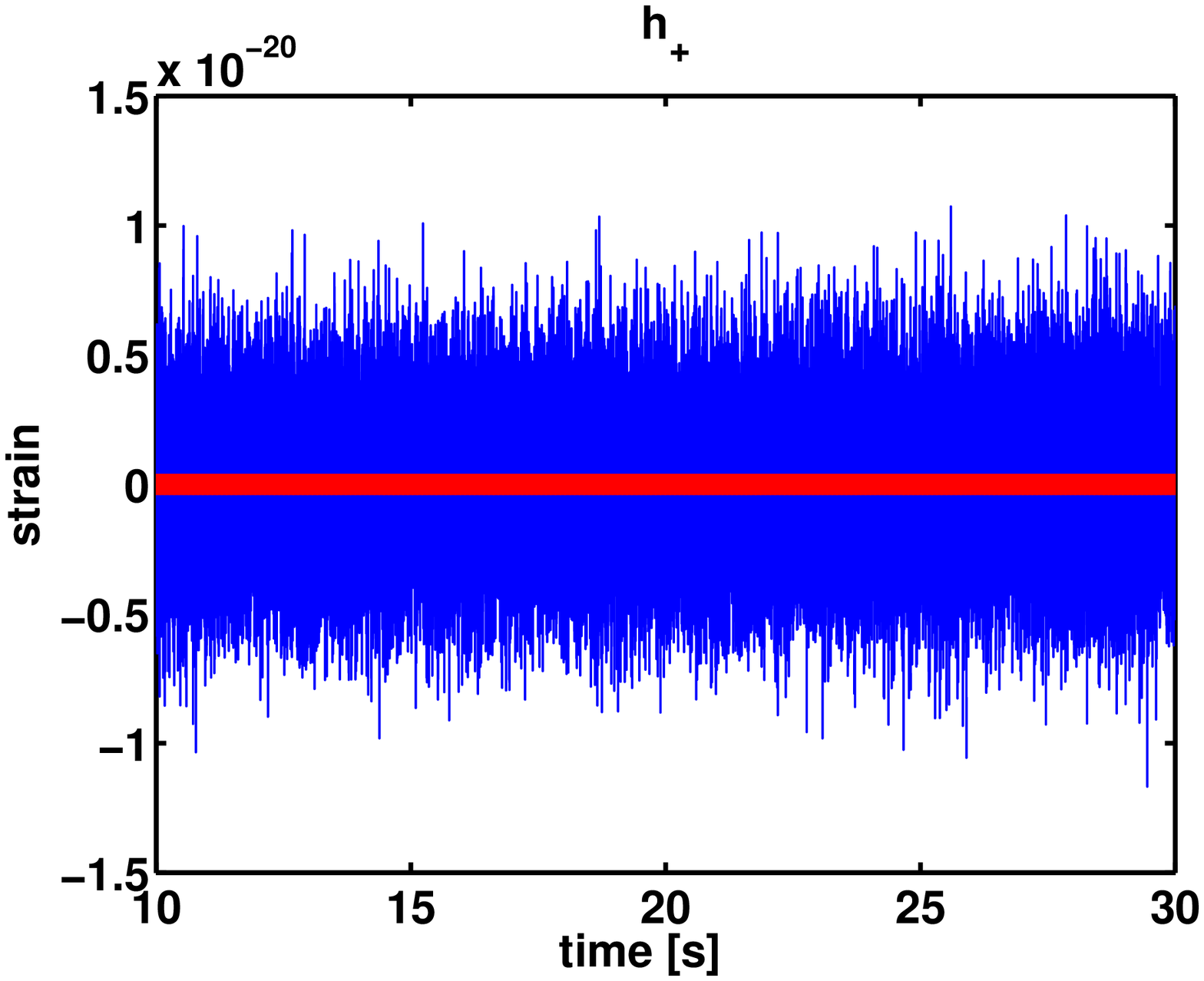}
\includegraphics[scale=0.3]{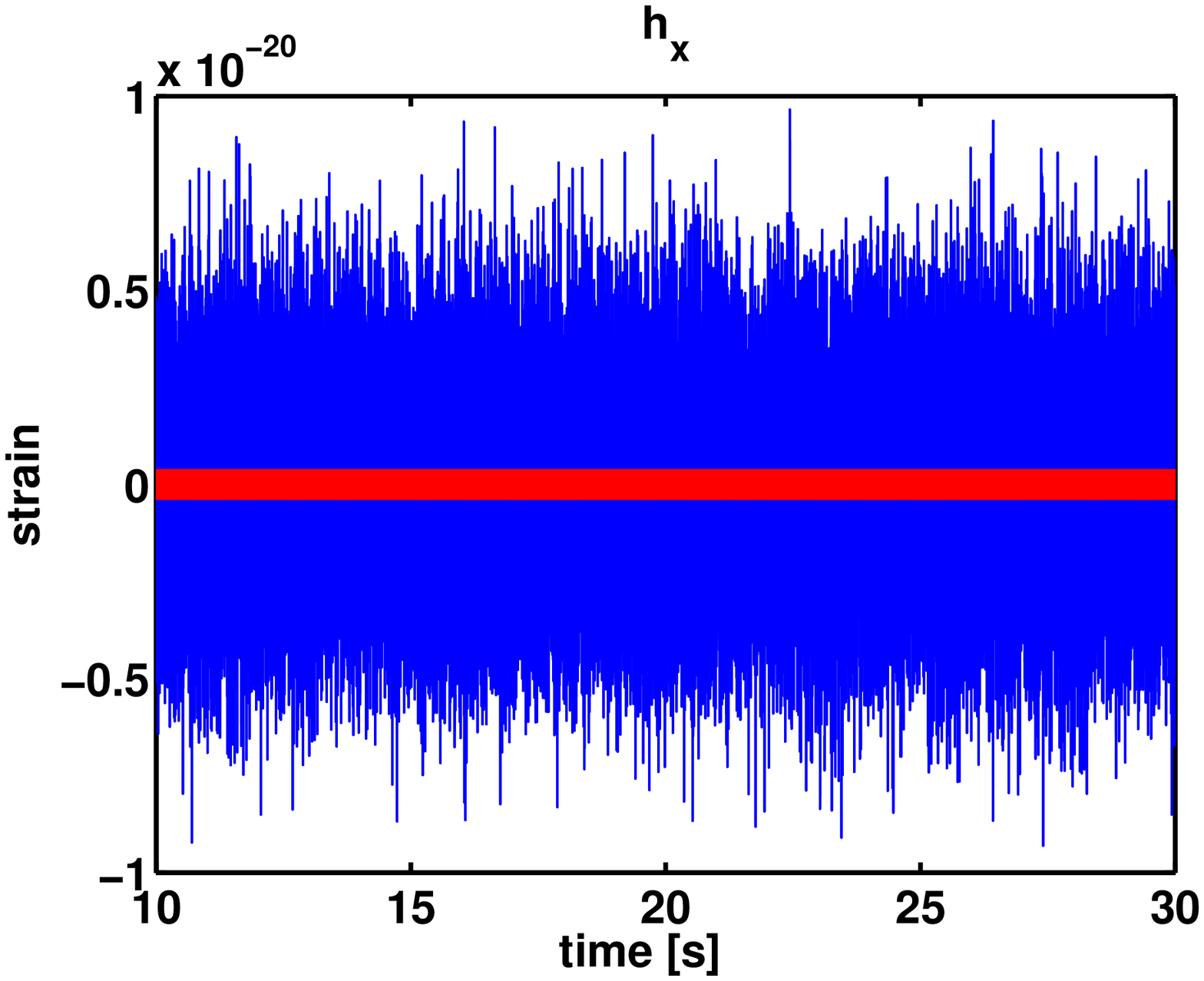}

\includegraphics[scale=0.3]{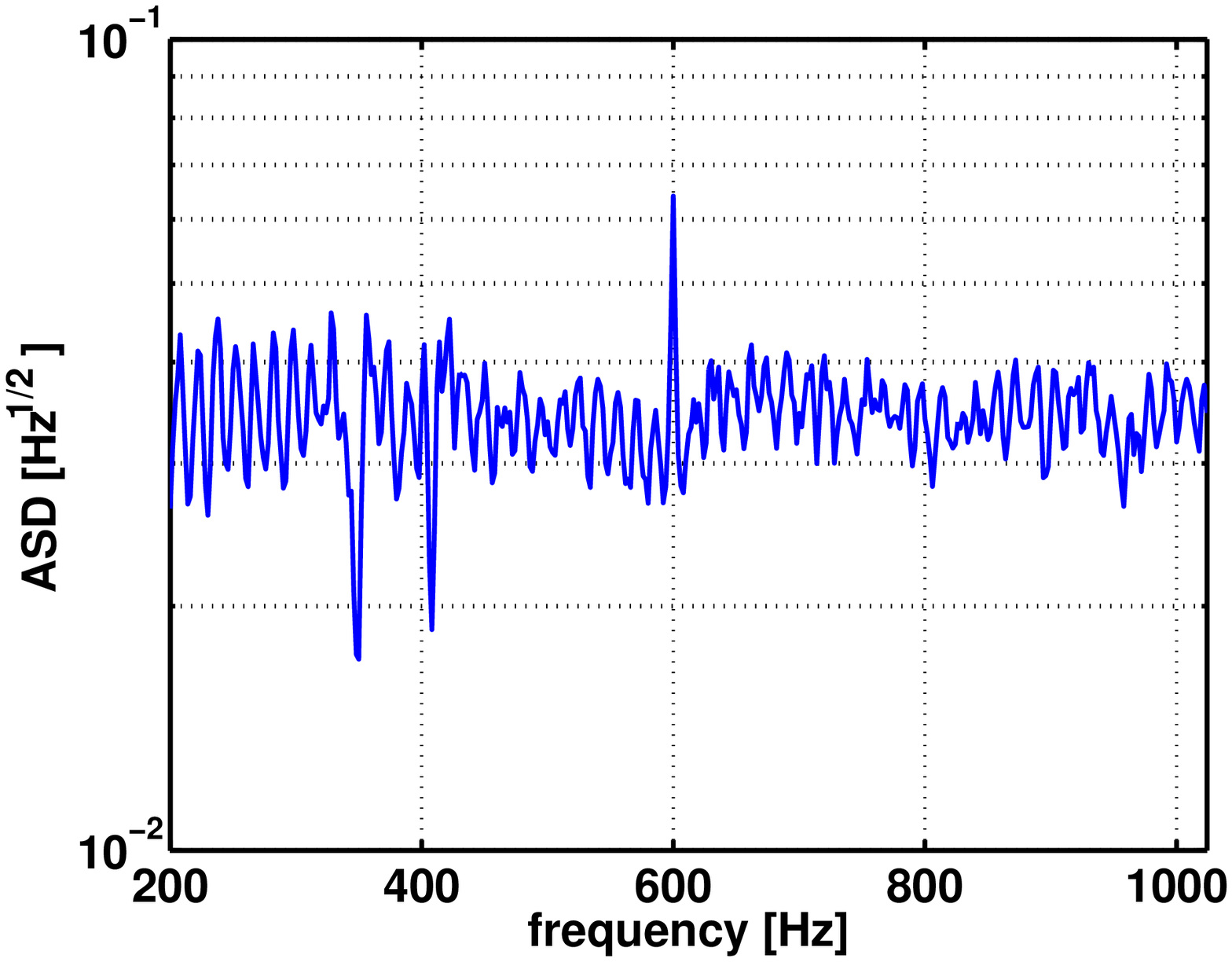}
\includegraphics[scale=0.3]{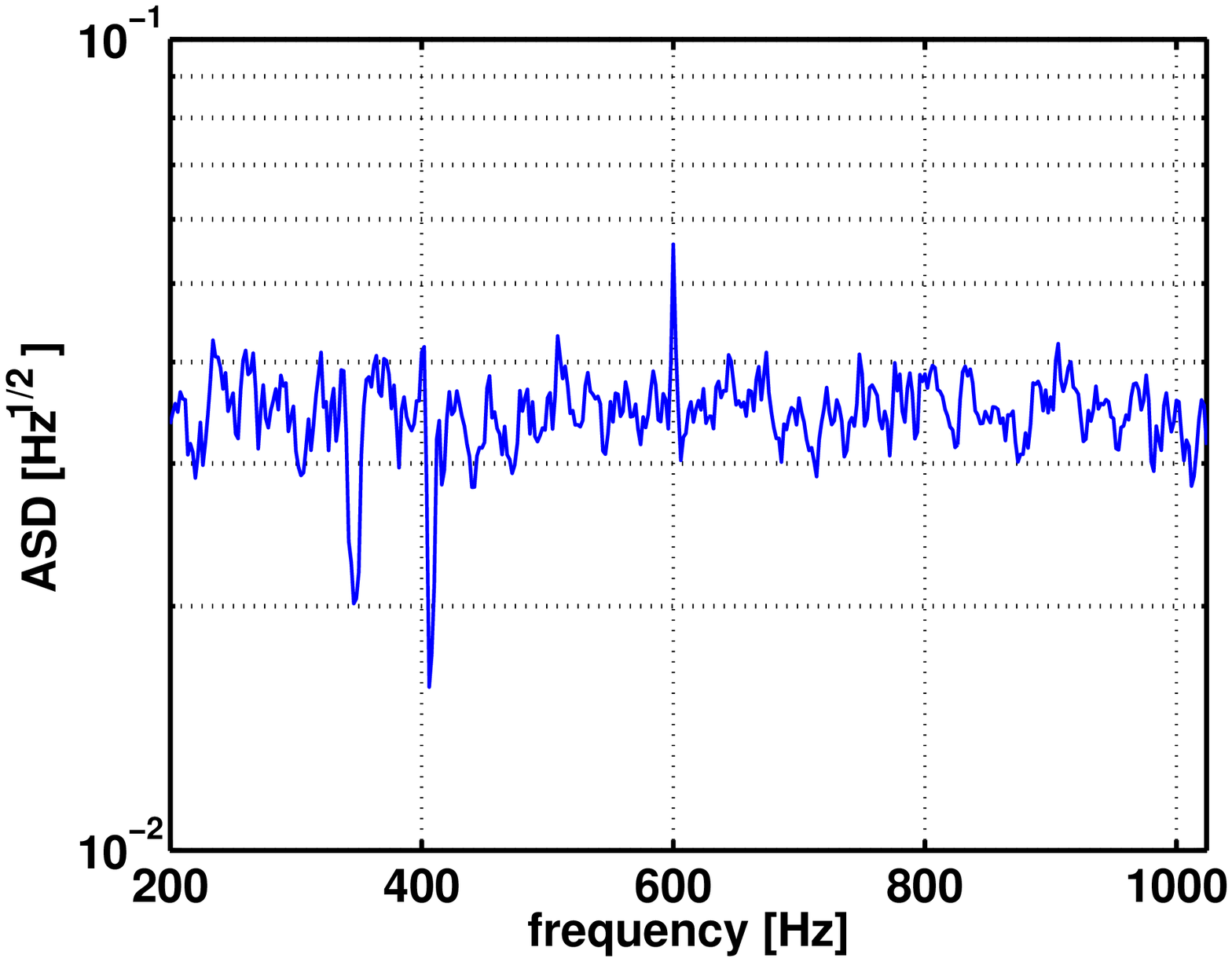}
\caption{Top plots show the reconstructed $h_+$ (left) and $h_{\times}$ (right), and red lines show injected $h_+$ and $h_{\times}$, of which frequency is 600 Hz and $h_{\rm rss}$ is $2.1\times10^{-21}{\rm Hz}^{-1/2}$. Bottom plots show the amplitude spectral density (ASD) of the reconstructed $h_+$ (left) and $h_{\times}$ (right). 
\label{fig:600hz}}
 \end{center}
\end{figure}

Finally we demonstrated the detection of a short duration monochromatic signal. The frequency of the injected signal was 600 Hz and the $h_{\rm rss}$ of the injected signal was $2.1\times10^{-21}{\rm Hz}^{-1/2}$. The top plots in figure~\ref{fig:600hz} are of the whitened reconstructed $h_+$ (left) and $h_\times$ (right). The blue lines are reconstructed $h_+$ (left) and $h_\times$ (right) and the red plots are the original injected signals. The lower plots are of amplitude spectral densities. The reconstructed $h_+$ and $h_\times$ of the signal can be seen clearly at 600 Hz in the plots.

%%%%%%%%%%%%%%%%%%%%
\section{Conclusion}
%%%%%%%%%%%%%%%%%%%%
We proposed the source tracking for Sco X-1 using the regularized coherent network analysis method {\tt RIDGE}. To demonstrate its performance, We considered the detection of unmodeled bursts, modeled bursts, and continuous sources. We applied an excess power method to the reconstructed $h_+$ and $h_\times$ for the detection of the unmodeled bursts. We also applied a matched filter method to the reconstructed waveforms for modeled signals. Finally, we reconstructed the short duration periodic signal, and applied Fourier transform to the reconstructed waveforms. These results show the source tracking approach works for a variety of signals.

%We find that even in its present simple form, the idea of source tracking is
%a viable one. An important issue is how to deal with GW signals which occur in the data
%and which have no connection with Sco X-1 as they may affect the
%reconstructed $h_+$ and $h_\times$ time series for the latter. 
%In a future paper, we shall show that the influence of GW signals from directions different from that of Sco X-1 can be eliminated by introducing a veto algorithm to see residuals for each detector data.
 
\ack{K.H. is supported by NASA grant NAG5-13396 to the Center for Gravitational Wave Astronomy at the University of Texas at Brownsville and NSF grant NSF-HRD0734800. SDM's work was supported by NSF grant PHY-0555842. S.Y. is supported by the Southeastern Louisiana University and NSF grant PHY 0653233.
 S.D. is supported by the Center for Gravitational Wave Physics at the Pennsylvania State
 University. The Center for Gravitational Wave Physics is funded by the National Science Foundation under
 cooperative agreement PHY 01-14375. T.S. is supported by a grant from the Office of Scholarly Research at Andrews University. This paper has been assigned LIGO Document Number LIGO-P080033-00-Z.
}
%%%%%%%%%%%%%%%%%%%%%
\section*{References}

\bibliographystyle{iopart-num}
\bibliography{scox1}

\providecommand{\newblock}{}
\begin{thebibliography}{10}
\expandafter\ifx\csname url\endcsname\relax
  \def\url#1{{\tt #1}}\fi
\expandafter\ifx\csname urlprefix\endcsname\relax\def\urlprefix{URL }\fi
\providecommand{\eprint}[2][]{\url{#2}}
% Bibliography created with iopart-num.bst, v1.0

\bibitem{Bildsten:1998}
Bildsten L 1998 {\em Astrophys. J. Lett.\/} {\bf 501} 89

\bibitem{Jernigan:2001}
Jernigan J 2001 {\em AIPC\/} {\bf 586} 805

\bibitem{1996A&AS..120C.641G}
Gruber D, Blanco P, Heindl W, Pelling M, Rothschild R and Hink P 1996 {\em
  Astron. Astrophys. Suppl.\/} {\bf 120} C641+

\bibitem{S2LSCpulsar:2007}
{Abbott B et al} 2007 {\em \PR D\/} {\bf 76} 082001

\bibitem{S4stochastic:2007}
{Abbott B et al} 2007 {\em \PR D\/} {\bf 76} 082003

\bibitem{Hayama:2007}
Hayama K, Mohanty S, Rakhmanov M and Desai S 2007 {\em \CQG\/} {\bf 24} S681

\bibitem{Mukherjee:2003}
Mukherjee S 2003 {\em \CQG\/} {\bf 20} S925

\bibitem{Chassande-Mottin:2005}
{The VIRGO collaboration} 2005 {\em \CQG\/} {\bf 22} S1189

\bibitem{Mohanty:2002}
Mohanty S 2002 {\em \CQG\/} {\bf 19} 1513

\bibitem{Rakhmanov:2006}
Rakhmanov M 2006 {\em \CQG\/} {\bf 23} S673

\bibitem{Klimenko+etal:2005}
Klimenko S, Mohanty S, Rakhmanov M and Mitselmakher G 2005 {\em \PR D\/} {\bf
  72} 122002

\bibitem{Mohanty+etal:2006}
Mohanty S, Rakhmanov M, Klimenko S and Mitselmakher G 2006 {\em \CQG\/} {\bf
  23} 4799

\bibitem{BradshawFomalontGeldzahler:1999}
Bradshaw C, Fomalont E and Geldzahler B 1999 {\em Astrophys. J.\/} {\bf 512}
  L121

\bibitem{Hayama:2008}
Hayama K, Mohanty S, Rakhmanov M, Desai S and Summerscales T 2008 {\em appear
  in Journal of Physics: Conference Series\/}

\bibitem{LIGOdesignsens}
Barish B and Weiss R 1999 {\em Phys. Today\/} {\bf 52} 44--50

\bibitem{VIRGOdesignsens}
Punturo M 1999 {\em {\rm VIRGO Report No. VIR-NOT-PER-1390-51}\/}

\bibitem{Anderson:2001}
Anderson W, Brady P, Creighton J and Flanagan {\'{E}} 2001 {\em \PR D\/} {\bf
  63} 042003

\end{thebibliography}

\end{document}